\documentstyle[twoside,fleqn,espcrc2,epsfig]{article}

\newcommand{\ba}{\begin{eqnarray}}
\newcommand{\ea}{\end{eqnarray}}
\newcommand{\be}{\begin{equation}}
\newcommand{\ee}{\end{equation}}
\newcommand{\lf}{\left}
\newcommand{\rr}{\right}

\newcommand{\ZV}{Z_V^{\rm eff}}

\font\tenmsb=msbm10 scaled\magstep1
\font\sevenmsb=msbm7 scaled\magstep1
\font\fivemsb=msbm5 scaled\magstep1
\newfam\msbfam
\textfont\msbfam=\tenmsb
\scriptfont\msbfam=\sevenmsb
\scriptscriptfont\msbfam=\fivemsb

\newcommand{\order}[1]{{\mathcal O}(#1)}

\newcommand{\AmS}{{\protect\the\textfont2
  A\kern-.1667em\lower.5ex\hbox{M}\kern-.125emS}}

\title{Semi-leptonic decays of heavy mesons}

\author{C.M. Maynard\address{Department of Physics and Astronomy, 
        The University of Edinburgh, \\ 
        Edinburgh, EH9 3JZ, Scotland, UK}\\
        UKQCD Collaboration}
       
\begin{document}

\begin{abstract}
We present results of a lattice computation of the matrix elements of 
the vector currents which are relevant for the semileptonic decays of 
$D \rightarrow K$, $D \rightarrow \pi$ and $B \rightarrow \pi$. 
The computations are performed in the quenched approximation on a  
$24^3\times 48$ lattice at $\beta=6.2$, using an $\order{a}$
non-perturbatively improved fermionic action.
\end{abstract}

\maketitle

\section{INTRODUCTION}
The matrix elements of heavy meson decays, particularly B mesons, are 
important for the determination of Cabibbo-Kobayashi-Maskawa matrix 
elements. We present preliminary results of a lattice study of vector 
current matrix elements for heavy-to-light semi-leptonic transitions.

\section{SIMULATION DETAILS}
We used the Wilson action to generate 216 $SU(3)$ gauge configurations on
a $24^3\times48$ lattice at $\beta=6.2$. The fermionic degrees of freedom
were computed using an ${\mathcal{O}}(a)$ improved clover action with a
non-perturbative value for $C_{SW}$ \cite{Alpha}.  The inverse lattice 
spacing, set by the $\rho$ mass is $a^{-1}=2.64$ Gev. Four values of the 
heavy quark hopping parameter were used: 0.1200, 0.1233,
0.1266, 0.1299, where 0.1233 roughly corresponds to the charm quark mass,
and all 4 are used to extrapolate to the b quark mass region.
Three values of the light quark hopping parameter were used: 0.1346, 
0.1351, 0.1353. The two heaviest were used 
for the spectator quark. The heavy quarks were smeared with novel gauge 
invariant functions 
and the light quarks fuzzed. 

We obtain the form factors from heavy-to-light three-point correlation
functions, dividing by the appropriate two-point functions.  We place
the operator for the heavy-light pseudo-scalar at $t=20$, 
rather than the midpoint of the lattice.  This allows us to estimate the
size of the systematic error coming from different time orderings and 
excited states. The tensor mixing with the vector current is included.

For $B \rightarrow \pi$ decay, results are only available for both 
initial ($\vec{p}$) and final ($\vec{k}$) states having zero momentum. 
For the $D$ decays, a range of momentum values were used to allow 
comparison with pole dominance models of the form factors.

\section{DECAYS OF D MESONS}
The matrix elements for D to K can be parameterised in terms of two 
form factors \cite{Bow95,pole},
\ba
    \left\langle K,\vec{k} \left | V^{\mu}\right |
        D,\vec{p} \right \rangle  & = &
    \left ( p + k - q \Delta_{m^2} \right )^{\mu}f_+(q^2) \nonumber \\ 
     & + & q^{\mu}\Delta_{m^2} f_0(q^2) 
  \label{eqn:Fpf0}
\ea
where
\ba
  \Delta_{m^2} = \frac{m^2_{K}-m^2_D}{q^2} & {\rm{and}} & q=k-p 
	\nonumber 
\ea
The form factors were measured for 6 different values of momentum 
transfer, $|p|\rightarrow |k|$: $0\rightarrow 0$, $0\rightarrow 1$, 
$1\rightarrow 1$, $1\rightarrow 0$, $1\rightarrow 1_{\perp}$, 
$1\rightarrow 1_{\gets}$, in units of $\pi/12a$.
Equivalent momentum channels are averaged over whenever possible 
to reduce statistical errors.

\subsection{Chiral Extrapolations}
\begin{figure}[hbt]
\vspace{-1.5truecm}
\begin{center}
\epsfig{figure=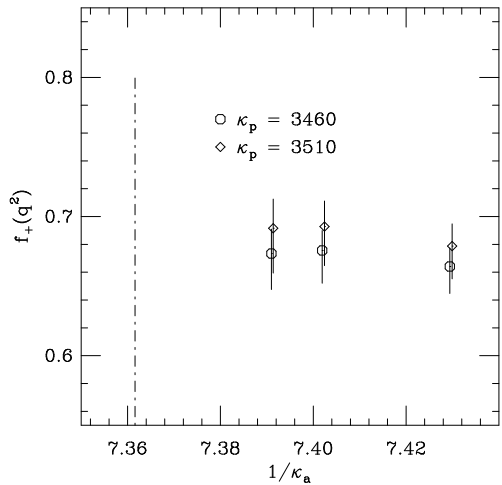,height=7cm,width=7cm} 
\vspace{-1.0truecm}
\caption{\em Chiral behaviour of $f^+$ for $\vec{p}=0$, $\vec{k}=(1,0,0)$  
	momentum channel.}
\label{fig:chiral}
\end{center}
\end{figure}

The extrapolation of the form factors and meson masses to physical values
of quark masses proceeds as follows.  For each momentum channel the six 
light kappa combinations were fitted to the following functional form 
\cite{Bow95}:
\ba
  &F(\kappa_a,\kappa_p)=\alpha 
   +\beta\lf(\frac{1}{\kappa_p}-\frac{1}{\kappa_c}\rr)+& \nonumber \\
 &\gamma\lf(\frac{1}{\kappa_p}+\frac{1}{\kappa_a}-\frac{2}{\kappa_c}
      \rr)^{\frac{1}{2}}
  +\delta\lf(\frac{1}{\kappa_p}+\frac{1}{\kappa_a}-\frac{2}{\kappa_c}\rr)&
\ea
where $\kappa_p$ refers to the passive or spectator quark and $\kappa_a$ 
to the active light quark at the $W$ vertex. Figure \ref{fig:chiral} 
shows the $f^+$ form factor for the six light kappa combinations.

\vspace{-1.0truecm}

The four-momentum transfer depends on the masses
of the states.  The meson masses were extrapolated using the 
PCAC relation and the momentum transfer  calculated using the 
dispersion relation at the extrapolated masses. We used the values of
$\kappa_{crit}=0.13844$ and $\kappa_{strange}=0.13476$.

\subsection{Pole Dominance Models}
Pole dominance models \cite{pole} suggest the following dependence of the
form factors on $q^2$
\ba
f^+(q^2)=\frac{f^+(0)}{1-q^2/m^2_{1^-}}&{\rm \&}&
    f^0(q^2)=\frac{f^0(0)}{1-q^2/m^2_{0^+}} \nonumber\\ 
    {\rm with} &&f^+(0)=f^0(0)  
\ea

\begin{figure}[hbt]
\vspace{-2.0truecm}
\begin{center}
\epsfig{figure=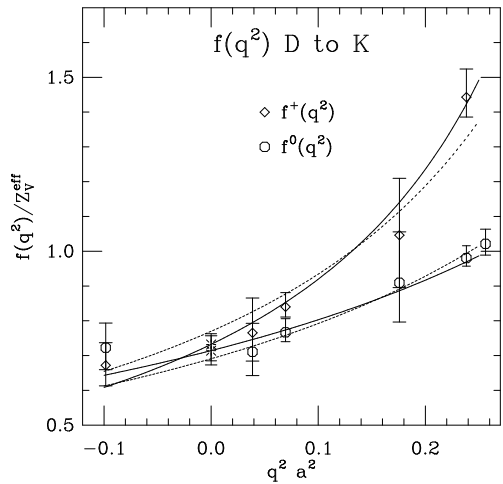,height=7cm,width=7cm} 
\vspace{-1.0truecm}
\caption{\em Pole dominance fits to form factors. The points at $q^2=0$ are 
	the interpolated data.}
\label{fig:pole}
\end{center}
\end{figure}

Figure \ref{fig:pole} shows the pole mass fits to the data.  The upper
curves are $f^+$ and the lower $f^0$.  The solid lines are a two 
parameter fit (fit A Table \ref{tab:pole}) for the form factor at zero
momentum and the pole mass. The dashed lines (fit B table \ref{tab:pole})
are a one parameter fit to the form factors, with the pole mass fixed 
to the two-point correlation function value.
\begin{table}[hbt]
\vspace{-1.0truecm}
\caption{\label{tab:pole}Pole Mass fits (masses in lattice units)}
\begin{tabular}{|c||cc|cc|}\hline
fit&$f^+(0)$&$m^{cs}_{1^-}$&$f^0(0)$&$m^{cs}_{0^+}$ \\ \hline\hline
A&$0.732^{+31}_{-47}$&$0.701^{+24}_{-24}$&$0.714^{+43}_{-41}$&
	$0.952^{+71}_{-72}$ \\ \hline
B&$0.770^{+24}_{-36}$&$0.754^{+2}_{-1}$ &$0.691^{+11}_{-21}$ &
	$0.884^{+12}_{-10}$ \\ \hline
\end{tabular}
\end{table}

The data show good agreement with pole dominance models.  For fit A, 
there is good agreement between $f^+(0)$ and $f^0(0)$, and the masses
of the poles agree with the two-point masses (the masses in fit B). For fit B,
the agreement between the vector and scalar form factors is not so good 
but still reasonable.

All the results quoted above are for $f^n(0)/\ZV$
The renormalisation constant, $\ZV$
can be evaluated in the Alpha scheme \cite{Alpha}, for D to K,
\be 
  \ZV=Z_V(1+b_Vm_Qa)=1.018
\ee

For D to $\pi$ the chiral extrapolations
are much harder, principally because we have to extrapolate a long way in
$q^2$ from the data. As a result the data for D to $\pi$ is not
as good and the pole mass dominance models do not work as well. More work
is needed on the chiral extrapolations.

\section{DECAYS OF B MESONS}
We have analysed the semileptonic decay $B \to \pi$ at zero recoil, i.e.
for $\vec{p}=\vec{k}=0$ that is $f^0(q^2_{\rm max})$. The form
factor is scaled to the $B$ mass by the following function, 
\cite{Theta}
\be
  \theta=\lf( \frac{\alpha_s(M)}{\alpha_s(M_B)} \rr )^{\frac{2}{\beta_0}}
\ee
where $\beta_0=11$ in the quenched approximation, and $\Lambda_{\rm QCD}
=295$Mev. The form factor is renormalised by $\ZV$ and then extrapolated
in $\frac{1}{M}$ with the following form,
\be
  \ZV f^0(q^2_{\rm max})\theta(M)\sqrt{M}=\eta + \frac{\gamma}{M} +
	\frac{\delta}{M^2}
\ee

\subsection{Pseudoscalar Decay Constant}
The mixing with pseudoscalar density is included.  The chiral 
extrapolation is linear in the ratio $f_B/f_{\pi}$.  Again we 
have the explicit mass dependence in the renormalisation constant.
Defining the ratio
\be
 Z_A^{\rm r}=\frac{Z_A(1+b_Aam_Q)}{Z_A(1+b_Aam_q)}
\ee
However $b_A$ is not known non-perturbatively, so we take a perturbative 
estimate \cite{Harm:lat97} $b_A=1.147$.  The ratio $f_B/f_{\pi}$ 
is extrapolated in $\frac{1}{M}$ in a similar fashion to $f^0$ except
renormalised by $Z_A^{\rm r}$.

\subsection{The Soft Pion Relation}
\vspace{-1.0truecm}
\begin{figure}[hbt]
\begin{center}
\epsfig{figure=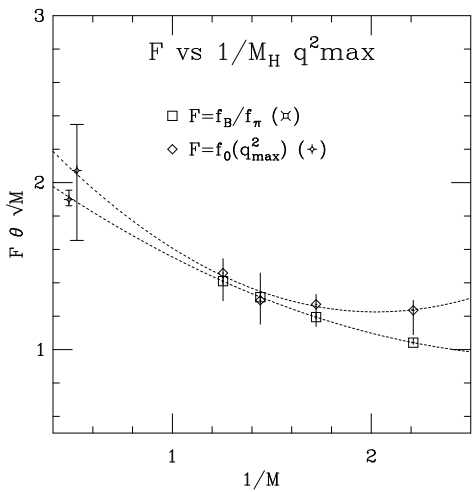,height=7cm,width=7cm} 
\vspace{-1.0truecm}
\caption{\em Heavy mass dependence of $f^0$ and ${f_B}/{f_{\pi}}$. The
	end points are the extrapolated data, offset from the B mass
	for clarity.}
\label{fig:b2pi}
\end{center}
\vspace{-1.0truecm}
\end{figure}
Figure \ref{fig:b2pi} shows the extrapolation of $f^0$ overlaid with that
of $f_B/f_{\pi}$.  It is clear that the soft pion theorem is 
satisfied, ie 
\be 
  f^0(q^2_{\rm max})=\frac{f_B}{f_{\pi}}
\ee
with
\ba
f^0(q^2_{\rm max})=1.47^{+20}_{-30} &{\rm and}&
	\frac{f_B}{f_{\pi}}=1.34^{+4}_{-3}
\ea
Two papers presented at Lattice 97
\cite{Tomi,Mats} found significant deviations from the soft pion 
relation. However there are several differences in the simulations.
Firstly, this work includes the renormalisation constants. Secondly, this
simulation is at much smaller lattice spacing. Thirdly, there are 
differences in the actions used.  This work uses a non-perturbatively 
improved action, \cite{Tomi} uses the Tadpole improved FNAL formalism for
heavy quarks and \cite{Mats} uses a NRQCD action.

\section{CONCLUSIONS}
We find that our results for D to K decays are in agreement with pole 
dominance models. For B to $\pi$ we find the soft pion relation 
satisfied. 

I would like to acknowledge the support of a PPARC studentship, and 
EPSRC grant GR/K41663 and PPARC grant GR/L29927.


\begin{thebibliography}{9}
\bibitem{Alpha} M. L\"{u}scher, S. Sint, R, Sommer, and P. Weisz, Nucl.Phys.B
	{\bf 478} 365 (1996)
\bibitem{Bow95} UKQCD Collaboration, K.C. Bowler {\em et al} Phys.Rev.D 
	{\bf 51} 4905 (1995)
\bibitem{pole} M. Bauer B. Stech \& M. Wirbel, Z.Phys.C{\bf 29}, 627 
    (1985); {\bf 34}, 103 (1987); M. Bauer \& M.Wirbel, {\em ibid}.
    {\bf 42}, 671 (1989)    
\bibitem{Theta} UKQCD Collaboration, J.M. Flynn {\em et al} Nucl.Phys.B
	{\bf 461} 327 (1996); D.R. Burford {\em et al} Nucl.Phys.B
	{\bf 447} 425 (1995)
\bibitem{Harm:lat97} H. Wittig Nucl.Phys.B(Proc.Suppl.) 63A-C 47 (1998)
\bibitem{Tomi} S. Tominaga {\em et al} Nucl.Phys.B(Proc.Suppl.) 63A-C 380
	(1998)
\bibitem{Mats} H. Matsufuru {\em et al} Nucl.Phys.B(Proc.Suppl.) 63A-C 
	368 (1998)
\end{thebibliography}
\end{document}